\documentclass[useAMS,usenatbib,a4paper]{aa} 
\usepackage[varg]{txfonts}
\usepackage{url}
\usepackage{units}
\usepackage{graphicx}
\usepackage{amssymb}

\newcommand\psrj{PSR~J2219+4754}

\bibpunct{(}{)}{;}{a}{}{,} 

\newcommand{\DMunits}{cm$^{-3}$pc}

\begin{document}

\title{First detection of
  frequency-dependent, time-variable dispersion measures}
\author{
	J.~Y.~Donner\inst{1,2}
	\and
	J.~P.~W.~Verbiest\inst{1,2}
	\and
	C.~Tiburzi\inst{2,1}
	\and
	S.~Os\l owski\inst{3,1,2}
	\and
	D.~Michilli\inst{4,5}
	\and
	M.~Serylak\inst{6,7}
	\and
	J.~M.~Anderson\inst{8}
	\and
	A.~Horneffer\inst{2}
	\and
	M.~Kramer\inst{2,9}
	\and
	J.-M.~Grießmeier\inst{10,11}
	\and
	J.~K\"unsem\"oller\inst{1}
	\and
	J.~W.~T.~Hessels\inst{5,4}
	\and
	M.~Hoeft\inst{12}
	\and
	A.~Miskolczi\inst{13}
}
\institute{
	Fakult\"at f\"ur Physik, Universit\"at Bielefeld, Postfach 100131,
	33501 Bielefeld, Germany 
	\and Max-Planck-Institut f\"ur Radioastronomie, Auf dem H\"ugel 69,
	53121 Bonn, Germany
	\and Centre for Astrophysics and Supercomputing, Swinburne
	University of Technology, P.O.~Box 218, Hawthorn, Victoria 3122,
	Australia
	\and Anton Pannekoek Institute for Astronomy, University of Amsterdam,
	Science Park 904, 1098 XH, Amsterdam, The Netherlands
	\and ASTRON, the Netherlands Institute for Radio Astronomy, Postbus 2,
	7990 AA, Dwingeloo, The Netherlands
	\and SKA South Africa, The Park, Park Road, Pinelands 7405, South Africa
	\and Department of Physics and Astronomy, University of the Western Cape,
	Private Bag X17, Bellville 7535, South Africa
	\and GFZ German Research Centre for Geosciences, Telegrafenberg, 14473 Potsdam, Germany
	\and	Jodrell Bank Centre for Astrophysics, School of Physics and Astronomy,
	The University of Manchester, Manchester M13 9PL, UK
	\and LPC2E - Universit\'{e} d'Orl\'{e}ans /  CNRS, 45071 Orl\'{e}ans cedex 2, France
	\and Station de Radioastronomie de Nan\c{c}ay, Observatoire de Paris,
	PSL Research University, CNRS, Univ. Orl\'{e}ans, OSUC, 18330 Nan\c{c}ay, France
	\and Thüringer Landessternwarte, Sternwarte 5, D-07778 Tautenburg, Germany
	\and	Astronomisches Institut, Ruhr-Universität Bochum, 44780 Bochum, Germany
	}
\date{Received 9 August 2018 / Accepted 6 February 2019}

\abstract
{High-precision pulsar-timing experiments are affected by temporal variations of the dispersion measure (DM), which are related to spatial variations in the interstellar electron content and the varying line of sight to the source.
Correcting for DM
  variations relies on the cold-plasma dispersion law which states
  that the dispersive delay varies with the squared inverse
  of the observing frequency. This may however give incorrect
  measurements if the probed electron content (and therefore the DM)
  varies with observing frequency, as is predicted theoretically
	due to the different refraction angles at different frequencies.
}
{We study small-scale density variations in the ionised
  interstellar medium. These structures may lead to
  frequency-dependent DMs in pulsar signals. Such an effect could
  inhibit the use of lower-frequency pulsar observations as tools to
  correct time-variable interstellar dispersion in higher-frequency
  pulsar-timing data. }
{We used high-cadence, low-frequency observations with three stations
  from the German LOng-Wavelength (GLOW) consortium, which are part of
  the LOw-Frequency ARray (LOFAR). Specifically, 3.5 years of weekly
  observations of \psrj{} are presented.}
{We present the first detection of frequency-dependent DMs towards any
  interstellar object and a precise multi-year time-series of
  the time- and frequency-dependence of the measured DMs. The observed
  DM variability is significant and may be caused by extreme
  scattering events. Potential causes
  for frequency-dependent DMs are quantified and evaluated.}
{We conclude that frequency-dependence of DMs has been reliably detected
  and is indeed caused by small-scale (up to 10s of AUs) but steep density variations in
  the interstellar electron content. We find that long-term trends in
  DM variability equally affect DMs measured at both ends of our
  frequency band and hence the negative impact on long-term
  high-precision timing projects is expected to be limited.
}

\maketitle

\section{Introduction}

Pulsars \citep[first discovered by][]{hbp+68} are highly magnetised, rapidly rotating neutron
stars, the remnants of massive stars that ended their life in a supernova.
Generally it is thought that pulsars emit beams of radiation
at their magnetic poles due to
magnetospheric effects that are not fully understood
\citep[e.g.][]{kja+15}. The magnetic and spin axes of pulsars are generally
not aligned, which causes the emission beams to sweep around in space
as the neutron star rotates. If one or both of the emission beams cross
the line of sight towards Earth during the rotation, regular pulses of radiation
can be detected, in which case the neutron star is called a pulsar.

The pulsed nature of the emission received from pulsars enables unique
and highly precise measurements of the electron density in
the ionised interstellar medium (IISM). This is due to the
frequency-dependent propagation speed of electromagnetic radiation in an
ionised medium, a phenomenon termed
dispersion. Specifically, the additional travel time for a wave
at frequency $\nu$ when compared to a wave at infinite frequency, is
approximated as \citep{lk05}:
\begin{equation}
	\label{eq:t_DM}
	\Delta t = D \frac{\rm DM}{\nu^2}
\end{equation}
where $D \simeq 4.149\times 10^3$\,MHz$^2$\,pc$^{-1}$\,cm$^3$\,s  is the dispersion
constant\footnote{Often the inverse is defined in the literature.} and
the `dispersion measure' DM (expressed in pc/cm$^3$) is defined as:
\begin{equation}
	\label{eq:DM}
	{\rm DM} = \int_0^dn_{\rm e}{\rm d}l,
\end{equation}
where $d$ is the distance to the pulsar (expressed in pc) and $n_{\rm e}$
is the electron density (in cm$^{-3}$).

This approximation makes use of the fact that $\nu \gg \nu_p$ and $\nu \gg \nu_c$,
where \citep[see][]{lk05}
\begin{equation}
	\label{eq:nu_p}
	\nu_p = \sqrt{\frac{e^2n_e}{\pi m_e}}
		= 8.98\,\text{kHz}\cdot\sqrt{\frac{n_e}{\text{cm}^{-3}}} \;\;\text{and}
\end{equation}
\begin{equation}
	\label{eq:nu_c}
	\nu_c = \frac{eB}{2\pi m_e c} = 2.80\,\text{MHz}\cdot\frac{B}{1G}
\end{equation}
are the plasma and cyclotron frequency respectively (in cgs-units), with $e$ and $m_e$ the charge and mass of an electron, respectively, $c$ the speed of light in vacuum and $B$ the magnetic field strength (in Gauss).
First-order deviations from the dispersion law (Eq.~\ref{eq:t_DM}) were described by \cite{tzd68}:
\begin{equation}
	\label{eq:tzd68}
	t_1 - t_2 = D\ \text{DM}\ \left(\frac{1}{\nu_1^2} - \frac{1}{\nu_2^2}\right)\ (1 + T_1 + T_2),
\end{equation}
with
\begin{equation}
	\label{eq:T1}
	T_1 = 3\nu_p^2\ (\nu_1^2 + \nu_2^2)\ /\ 4\nu_1^2\nu_2^2 \;\;\text{and}
\end{equation}
\begin{equation}
	\label{eq:T2}
	T_2 = \pm 2\nu_c \cos\gamma\ (\nu_2^3 - \nu_1^3)\ /\ \nu_2\nu_1(\nu_2^2 - \nu_1^2).
\end{equation}
$T_1$ and $T_2$ are only dependent on the electron density and the magnetic field strength along the line of sight, respectively, as well as the observing frequencies in question. \cite{tzd68} could not find any evidence for a deviation from Eq.~\ref{eq:t_DM} in their dataset. More recently, \cite{hsh+12} has come to the same conclusion.

Since pulsars are typically high-velocity objects \citep{go70,ll94},
their lines of sight travel through the Galaxy with sufficient speed
that variations of DM in time (corresponding to spatial inhomogeneities in the IISM)
are regularly observed \citep[see, for
example,][]{rtd88}. However, such variations can only be
accurately measured and distinguished from other noise sources if the
fractional bandwidth of the observations is sufficiently large or if a
range of observing frequencies are available, even though they do
(possibly significantly) affect narrow-band or frequency-integrated
observations as well \citep{lsc+16}.

The importance of accurate measurements of time-variable DM values
lies in the main applications of pulsars. Due to their extremely high
rotational stability \citep[rivalling atomic clocks on the long term:][]{hcm+12},
pulsars have become one of the main tools with which to test a wide variety of
physics, from the equation of state at super-nuclear densities
\citep{lp16} to general relativity and a variety of alternative
theories of gravity \citep[and references therein]{wil14}. These
pulsar-timing\footnote{For an introduction to pulsar timing, see \cite{lk05}.}
 tests, however, typically take place at relatively high
frequencies -- generally around 1.4~GHz,
where for most pulsars a useful balance is found between the
brightness of the pulsar itself and the Galactic synchrotron
background noise; where RFI is relatively limited; and where high-quality
receiver systems are commonly available. The fact that the
interstellar dispersion (and hence the variations in interstellar
electron content) only has a limited impact at these frequencies, is
also in principle a positive aspect, as it
prevents corrupting effects from IISM turbulence in the pulsar-timing
data.

For long-term high-precision timing projects, however, this situation
may change since the power spectrum of the turbulent structures in the
IISM is steep, with significantly more power at the
larger scales \citep{ars95}. This implies that for the most precise
and longest-term pulsar-timing projects \citep[like the `pulsar timing
array' (PTA) projects which aim to detect gravitational waves, see][]{tib18} it may
not suffice to opt for these higher frequencies, since sooner or
later IISM turbulence may still become a problem. In their discussion
of PTA experiments with the Square Kilometre Array (SKA),
\citet{jhm+15} propose ways to mitigate or prevent corruption of
PTA data by variations in pulsar DMs. One approach uses low-frequency
data (specified as between $\sim$100 and $\sim$300~MHz) to
independently monitor the IISM and construct DM time series that can
be used to correct time variability in the DMs at higher frequencies
(which are more sensitive to the pulsar but less to the interstellar
dispersion).

While most of the observed DM variations to date have been interpreted
as being caused by the IISM's turbulence, a second source of
variability was identified in the so-called `extreme scattering
events' (ESEs).\footnote{While the term ESE has historically been used for the event only,
	we also use the term here to refer to the underlying ISM structures, as is now commonly done.}
ESEs were first discovered as rapid variations in AGN
flux densities \citep{fdjh87}, but were soon also detected in the flux
density, arrival times and DMs of pulsars \citep{cbl+93,mlc03,kcw+18}. A
variety of models and origins of ESEs has been proposed
\citep[see, e.g.][]{wal01a}, though commonly two prime models are used:
either the ESEs are treated as individual `lenses', that is, local
overdensities or self-contained clouds \citep[see,
  e.g.][]{rbc87,ww98,cbl+93}, or they are seen as part of the
larger-scale turbulent structure in the IISM \citep[as suggested
  by][]{fdj+94,cks+15}. One possible way to differentiate between
these two scenarios would be to probe the turbulence within an ESE,
as attempted by \citet{lfd+00}. As reviewed by \citet{bch+15}, the
origin and IISM role of ESEs are at present not fully understood.
However, their potential relation to scintillation arcs
\citep[higher-order interference in pulsar scintillation,
  see][]{smc+01} does suggest that filament-like structures \citep{bmg+10},
which may be part of the larger-scale IISM turbulence \citep{pk12,pl14},
could contribute to the solution. More recently, \citet{cks+15} found
that several ESEs that were observed in pulsar observations did appear
to be related to the IISM's Kolmogorov turbulence. In the context of
pulsar timing, understanding the prevalence, origin and nature of ESEs
is crucial before their impact on timing experiments (and their
potential mitigation) can be evaluated.

In correcting time-variable DM delays in high-precision pulsar timing
data (as discussed above), a major potential problem lies in the
possibility of frequency-dependent DMs, also known as `chromaticity'.
Such a phenomenon could be induced in two different ways.
As the dispersion law (Eq.~\ref{eq:t_DM}) is an approximation for the case
$\nu\gg\nu_p$ and $\nu\gg\nu_c$, it could be invalid in extreme cases
(e.g.\ low frequencies, large electron densities or magnetic field strengths).
In that case, assuming the dispersion law to be accurate would lead to
a different DM measured at different frequencies (see Eqs.~\ref{eq:tzd68} to \ref{eq:T2}).

Chromaticity could also be induced by the fact that
variations in electron density in the IISM do not only cause
time variations in the measured dispersion, they also induce
refraction of the radiation on several different spatial scales. This
refraction causes rays to not travel along a perfectly straight line
from the pulsar to the observer, but rather over some sort of `random
walk'. Since the strength of the refraction is frequency-dependent,
this also means that the photons we receive at different frequencies
traverse different parts of the IISM and therefore may sample regions with
different electron densities. In principle this would lead to
frequency-dependent measurements of DM in the case of inhomogeneous
media. As demonstrated by \citet{css16} both theoretically and through
simulations, the fact that the IISM volumes sampled by the radio waves
differ across frequencies effectively causes the DM time series
observed at low frequencies to be very similar to a low-pass-filtered
(or smoothed) version of the DM time series measured at higher
observing frequencies \citep[see Figures~3 and 4 of][]{css16}.

In practice, this effect has not been observed, although limits have been
placed using extremely wide ranges of frequencies \citep{hsh+12,pdr14}. An
observational test of this phenomenon would allow realistic tests of how such
DM chromaticity may affect the usefulness of
low-frequency DM time series for correcting higher-frequency
pulsar-timing data.

In this paper, we present high-cadence low-frequency observations of
\psrj, a slow pulsar discovered by \citet{th69} with
a DM of 43.5\,cm$^{-3}$\,pc. Observed with the LOw Frequency ARray (LOFAR)
telescope, it is one of the brightest sources in the northern sky
\citep[see][]{bkk+16}. Combined with the high fractional bandwidth of LOFAR,
this leads to a very high DM measurement precision \citep[see][]{vs18}.
Strong DM variations have previously been reported for this source by \citet{agmk05}.
Additionally, this pulsar is known to show profile-shape variations.
While \citet{ss94} concluded that the profile-shape variations they observed are intrinsic
to the pulsar, more recently, the analysis of our companion paper 
\citep{mhd+18} suggests an interstellar origin.

In Section~\ref{sec_obs} we describe the observations used
in our work, while Section~\ref{sec_analysis} explains the steps taken
in deriving the DM time series (and the detected frequency dependence
of the measured DM values). Section~\ref{sec_discussion} discusses the
nature of the DM variations observed and assesses possible
implications for pulsar timing. Section~\ref{sec_concl} concludes the
paper by summarising our main findings.

\section{Observations}\label{sec_obs}
Our analysis is based on data from three German stations of the
International LOFAR Telescope \citep[ILT,][]{vwg+13},
namely the stations in Effelsberg (telescope identifier DE601),
Tautenburg (DE603) and J\"ulich (DE605), between 25 February 2013 and
25 November 2016 (see Table~\ref{tab:obs}).
LOFAR is described in
detail by \citet{vwg+13} and some aspects of particular relevance to
pulsars are described more in-depth by \citet{sha+11}. In contrast to
the set-up described in those papers, the observations used in our
work were carried-out in a `stand-alone' mode, in which the
individual stations were disconnected from the ILT network and used
as independent telescopes. While in stand-alone mode,
the beamformed data were sent from the stations in Effelsberg,
Tautenburg and J\"ulich to the Max-Planck-Institut f\"ur
Radioastronomie (MPIfR) on dedicated, high-speed links, where
recording computers ran the dedicated LOFAR und MPIfR Pulsare
(\textsc{LuMP}\footnote{Publicly available at
  \url{https://github.com/AHorneffer/lump-lofar-und-mpifr-pulsare} and
  described on
  \url{https://deki.mpifr-bonn.mpg.de/Cooperations/LOFAR/Software/LuMP}.})
data-taking software, which formats and otherwise prepares the
beamformed pulsar data for subsequent (off-line but near-real-time)
phase-resolved averaging (commonly referred to as `folding') using
the \textsc{dspsr} software package \citep{vb11}. This produces data
cubes with resolution in frequency (195.3125\,kHz-wide channels), time
(10-sec sub-integrations), polarisation (four coherency products) and
rotational phase (1024 phase bins).

\begin{table*}
	\centering
	\caption{Summary of observations.
      Given are the telescope identifier; the number of
      observations $N_{\rm obs}$; the time span of the observations; the
      range of observation lengths and the median observation length;
      the bandwidth of the observations (which changed for DE601 and
      DE605 as discussed in Section~\ref{sec_obs}) and the
      geographical location of the stations.
	}\label{tab:obs}
	\begin{tabular}{c c c c c c c c}
		Telescope  & $N_{\rm{obs}}$ & Gregorian                & MJD            & \multicolumn{2}{c}{Observation length} & Bandwidth  & Location \\
		identifier &                & date range               & range          & range         & median                 & (MHz)      & \vspace{4pt} \\
     \hline \rule{0pt}{15pt}\unskip                                              
		DE601      & 17             & 24/05/2013 -- 08/01/2014 & 56436 -- 56665 & 7 -- 60\,min  & 28\,min                & 47.7, 95.3 & Effelsberg \\
		DE601      & 2*             & 19/05/2015 -- 20/05/2015 & 57161 -- 57162 & 9\,hrs, 8\,hrs&                        & 95.3       & Effelsberg \\
		DE603      & 10             & 12/02/2014 -- 03/05/2014 & 56700 -- 56780 & 7 -- 13\,min  & 13\,min                & 95.3       & Tautenburg \\
		DE605      & 119            & 07/03/2014 -- 25/11/2016 & 56723 -- 57717 & 2 -- 146\,min & 115\,min               & 95.3, 71.5 & J\"ulich \\
        \hline
	\end{tabular} \\
	\vspace{1ex}\raggedright\qquad
	*These observations were omitted from the analysis, and are used as reference template instead.
\end{table*}

A few changes to the set-up of this network were made over the years
during which these observations were taken. Specifically, in August 2013 a
new data-taking mode was deployed to allow a reduction in the
number of bits with which the data were recorded (reduced to eight bits
from the original 16 bits). Since the total data rate which the
recording computers and network links can keep up with limits the
total amount of data that can be recorded, this reduction in bits
enabled an increase in observing bandwidth. Consequently, the observing
bandwidth doubled from 47.7\,MHz to 95.3\,MHz starting in mid August
2013. This change in bandwidth also slightly changed the centre
frequency of the observations, which moved from 138.77\,MHz to
149.90\,MHz at that time. This implies a shift of the centre
frequency by an integer number (57) of frequency channels. (The last
observation with 47.7\,MHz of bandwidth in our data set was taken on
19 August 2013 while the first with 95.3\,MHz of bandwidth was
recorded on 27 August 2013.)

For technical reasons, we restricted the bandwidth of observing to
71.5\,MHz from February 2015 onwards. In order to minimise the impact of this bandwidth
reduction on the scientific quality of the data, the observed
bandwidth was kept centred on the most sensitive part of the bandpass,
thereby causing the centre frequency to shift slightly from
149.90\,MHz to 153.81\,MHz. This shift was again made by an integer
number of frequency channels (20 channels in this case), so that the frequencies
of individual channels remained constant over the entire dataset.

\section{Data analysis}\label{sec_analysis}
The data analysis has been carried out using the \textsc{psrchive}
\citep{hvm04,vdo12}, \textsc{tempo2} \citep{hem06} and
\textsc{coastguard} \citep{lkg+16} software packages, as detailed below.

\subsection{Pre-processing}
Before any of the other analysis steps were carried out, the data were
inspected for man-made radio-frequency interference (RFI) and any
affected channel-subintegration combinations were removed from the data using
the `Surgical' algorithm from the \textsc{clean.py} script,
which is part of the \textsc{coastguard} python package.\footnote{
	Publicly available at \url{https://github.com/plazar/coast_guard}.}
In addition to this, throughout the analysis the
original data from outlier points were also visually
inspected to ensure the absence of RFI. Where needed, RFI was
manually removed using the \textsc{psrchive} program \textsc{pazi}
upon which the processing was repeated. (The number of data points that
were reprocessed in this way was minimal.) Typically about 25\%
of our data were excised due to presence of RFI. This percentage
varied depending on the time of day of the observation (with higher
prevalence of RFI during the day and lower during the night) but no
significant difference was seen in the RFI fraction of the different
stations.

The data were calibrated in polarisation
following the methods outlined in \citet{nsk+15}, after which the coherency products
were combined to yield total intensity. Given the
limited length of the observations (typically shorter than two hours) and the
fact that the hour angle and elevation of all observations were highly similar
as the observations were always scheduled close to or across transit, the
calibration (and therefore also its imperfections)
did not significantly affect our analysis. Next, the total-intensity
profiles were averaged in time, resulting in a single, frequency and
phase-resolved pulse profile for every observation.

The data were not averaged in frequency, but in order to produce as
homogeneous a data set as possible, the data observed with 95.3\,MHz
of bandwidth were downsized to 71.5\,MHz instead, by cutting out the
edges of the band, where the sensitivity is low due to the presence of filters \citep{vwg+13}.
The full-bandwidth data remains available upon request for possible
follow-up investigations, when needed.

\subsection{Timing and DM time series}\label{ssec:Timing}
In order to determine the DM time series, we used the pulsar-timing
technique \citep[see][]{lk05}.
While an accurate timing model is in principle
not required for instantaneous measurements of DM, the time-averaging
of observations does improve if the pulsar timing model is
of good quality. Consequently we carried out an initial, straightforward
timing analysis based on a single analytic template profile
constructed of von Mises functions \citep[see, e.g.][]{js01} of the form:
\begin{equation}
	f(x) = A \cdot e^{\kappa\cdot(\cos 2\pi(x - \mu) - 1)}
\end{equation}
with $A$ being the amplitude of the component, $\kappa$ the so-called compactness,
and $\mu$ the pulse phase. One full rotation corresponds to one unit in $x$.
Around five of these functions were fitted to an arbitrary, fully
frequency-averaged high-S/N observation, using the \textsc{psrchive} program
\textsc{paas}. This analytic template was cross-correlated against the
profiles of each frequency channel of each observation \citep[using
  the standard method described by][]{tay92}, after which the
\textsc{tempo2} software package was used to fit for the
DM at each observing epoch. Subsequently these daily DM measurements
were held fixed in the timing model of \citet{hlk+04} while the entire
data set was used to fit the pulsar's spin period, spin period
derivative and position. This updated timing model was then used to
re-do the time-averaging of the observations, after which the process
was iterated until the timing model and DM time series converged.

After this initial timing analysis, the proper motion in the timing model was
updated from the values of \citet{las82} to those of \citet{mhd+18}
(which was not available at the start of our analysis), although this
had no significant impact on our results given the short time-span of our
observations. Since this pulsar exhibits large amounts of timing noise
\citep{hlk+04} and because our timing analysis fully ignored
frequency-dependent profile evolution and the temporal evolution of
the profile described in our companion paper \citep{mhd+18}, the
results of our timing analysis were not ideal and certainly not
predictive enough to warrant publication of the timing model. However,
the timing model thus obtained did succeed in phase-aligning our
observations to within $\sim 400\,\mu$s, that is, within a phase bin; and
much more precisely within any given observation. We therefore used
this timing model for the final time-averaging of our data.

In order to determine highly precise and reliable DM values, a more advanced
timing analysis was carried out. For this analysis we used a frequency-resolved
template. To create this template we combined two long observations taken with
DE601 on 19 May 2015 (MJD~57161), from 03:00 to 12:00 UTC and on 20 May 2015
(MJD~57162), from 00:00 to 07:50 UTC, for a total effective duration of 16.6
hours. This observation was averaged
in time and summed to total intensity, providing a frequency-resolved pulse
profile with a S/N a few times that of the typical observation. This template was
subsequently used as the phase reference for timing and was otherwise fully
omitted from our analysis. The pulse times-of-arrival (ToAs) were in this case
determined using the Fourier-Domain Monte-Carlo (FDM) approach\footnote{This algorithm
	is identical to that described by \citet{tay92}, except for the uncertainties.
	FDM uses either formal uncertainties or a Monte-Carlo simulation.
	We used the default formal uncertainties.}, as advised by
\citet{vlh+16}, on a channel-by-channel basis (i.e.\ resulting in up to 366
simultaneous ToAs per observation). Since the template profile is data-derived
and frequency-resolved, any static frequency dependence of the template shape
would not affect our analysis, as it is inherently taken into account. (Also,
any DM measurement derived from this analysis is by definition referred to the
DM incorporated in this template, thereby rendering an absolute DM measurement
impossible.)  However, time variability of the pulse profile shape \citep[as
  investigated by][]{mhd+18}, does have the potential to negatively impact the
reliability of our results. This is discussed in detail in
Section~\ref{DMvars_test}.

\begin{figure}
	\includegraphics[width=.5\textwidth]{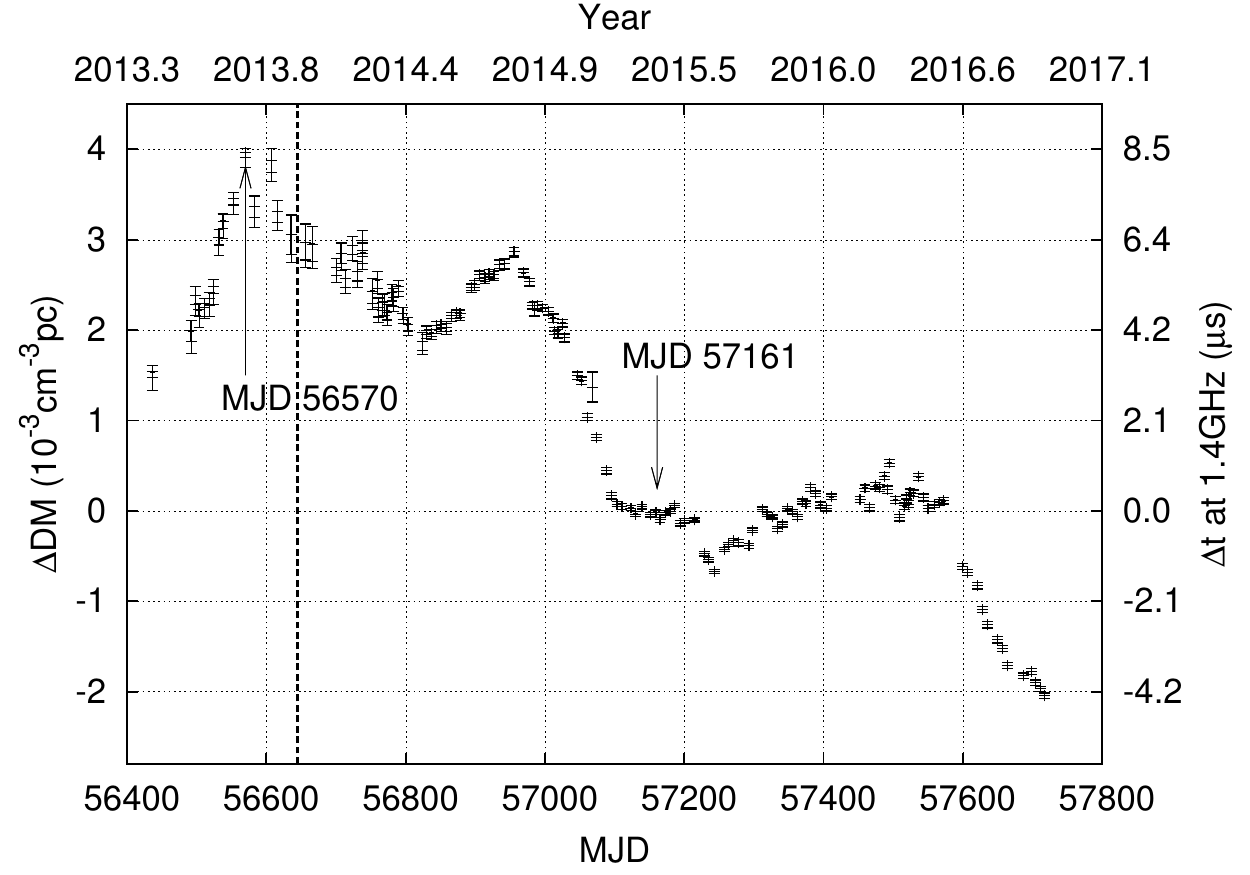}
	\caption{DM variations in the direction of \psrj. The
      vertical line indicates the time at which the last change to the
      centre frequency and bandwidth of our data occurred, i.e.\ all
      data to the right of the dashed line have identical bandwidth
      and centre frequency (except for slight deviations due to
      variations in RFI excision). The two arrows indicate the observations
      that were used to quantify the effect of scattering (see
      Section~\ref{DMvars_test}), with the latter of these two (MJD~57161) also
      being the standard template. A DM baseline of
      $43.48205\,$\DMunits\ has been subtracted. The additional error bars
      on the lower end of the pre-MJD~57000 data points indicate the
      expected impact of profile scattering, as described and
      quantified in Section~\ref{DMvars_test}.
	 The second y-axis indicates the corresponding dispersive delay at
	 an observing frequency of 1.4\,GHz.}
	\label{fig_DMvars}
\end{figure}

In this more advanced timing analysis we do not fit any time-dependent
timing-model parameter but exclusively for DM on an
observation-by-observation basis. Since these DM fits are based on a
set of simultaneous ToAs (i.e.\ those derived from the different
channels of the given observation), no time-dependent timing-model
parameters affect our results. An arbitrary phase offset is routinely
subtracted along with every fit, in order to prevent biases discussed
by \citet{kcs+13}.
The DM measurements from this analysis had a median uncertainty of
$3.7\times10^{-5}$\DMunits\ and their time series is presented in
Figure~\ref{fig_DMvars}, but before deeper consideration of these DM
values, some corrupting influences will be discussed below.

\subsection{Impact of pulse-shape variations on the DMs}
\label{DMvars_test}
\begin{figure}
	\includegraphics[width=.5\textwidth]{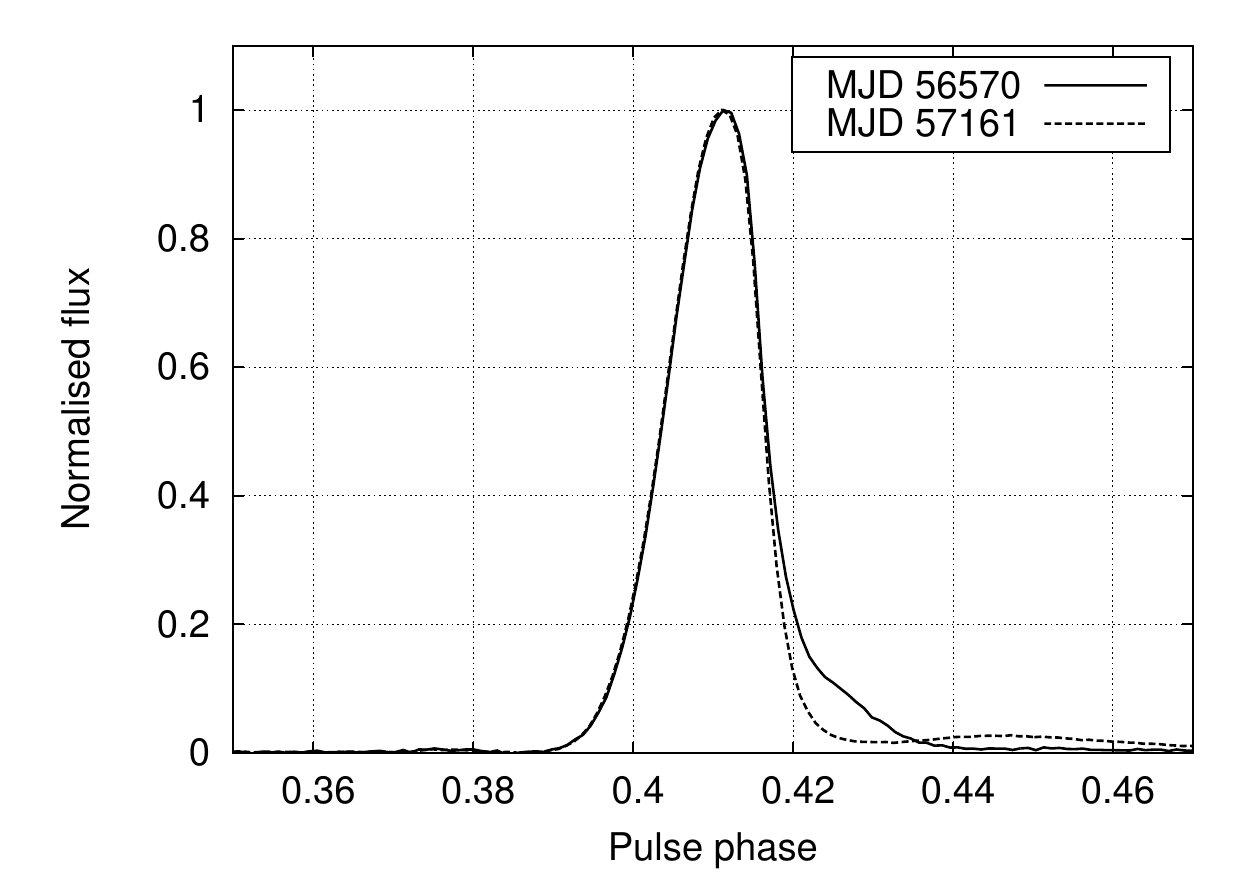}
	\caption{Peak-aligned pulse profiles for a scattered observation on
      MJD~56570 and an unscattered observation on MJD~57161, which is part
      of the standard template. The profiles were aligned to have the leading
      edge and peak (which have identical shape at both dates) to align; and the
      pulses shown are integrated over the full bandwidth of 71.5\,MHz.}
	\label{fig_shapeComparison}
\end{figure}

As mentioned in the previous paragraph, time variations in the shape
of the pulse profile could corrupt our measured DM values,
particularly if this time variability is also frequency
dependent.
The profile-shape evolution discussed in our companion paper \citep{mhd+18} is
of particular concern since it is known to affect our data
and has been shown to be frequency dependent. Specifically,
\citet{mhd+18} report time-variable scattering that shows up as
additional pulsed components at and near the trailing edge of the
pulse profile. As scattering is frequency dependent, so are these
components, being more pronounced at lower frequencies than at higher
frequencies. The net effect of such additional profile components on
our DM measurements would be to delay the measured ToAs; and since the
additional components are more pronounced at lower frequencies, this
would lead to an overestimate of the DM.
Figure~\ref{fig_shapeComparison} shows the worst-case profile-shape
variability\footnote{As quantified by the goodness-of-fit
  of the template cross-correlation during the timing with the FDM
  algorithm described in the previous section. While the scattering
  strength does appear to be marginally stronger at earlier dates, the
  narrower bandwidth of the earlier observations cause the
  profile-shape difference to be partly covered up by increased levels
  of radiometer noise.} present in our data set, comparing the pulse
profile at MJD~56570, when the scattering components are most pronounced, to
the standard template from MJD~57161 when the scattering is minimal,
but some reflected `echo' images of the
main pulse are faintly visible at longer lags \citep[see][for a full
  discussion of these pulse-shape variations]{mhd+18}.

To quantify the impact of the scattered power in the trailing edge
of the pulse profile shown in
Figure~\ref{fig_shapeComparison}, an analytic model of the
(unscattered) template profile was extended with
extra power in the trailing edge to match the one
shown in the observation of MJD~56570.
By timing both of these analytic pulse
models (the one with additional power in the trailing edge and the one without)
and repeating the process at three different frequencies, we
could directly determine the impact the scattering has on the ToAs
at different frequencies.

Fitting for the DM for the scattered and unscattered profiles returns a 
difference of $\Delta \text{DM} = 1.5(4)\times10^{-4}\,$\DMunits. We note
this DM difference is the maximum offset induced by profile-shape
changes.
A similar test for the reflected `echo' images visible at later
MJDs lead to an insignificant impact on the DM.
In comparison to the DM variations and DM measurement precision shown
in Figure~\ref{fig_DMvars}, we note that enhanced scattering by this
amount does slightly alter the results, but does not fundamentally
change the shape of the (much stronger) DM variations we identified. 

\begin{figure}
	\centering
	\includegraphics[width=0.5\textwidth]{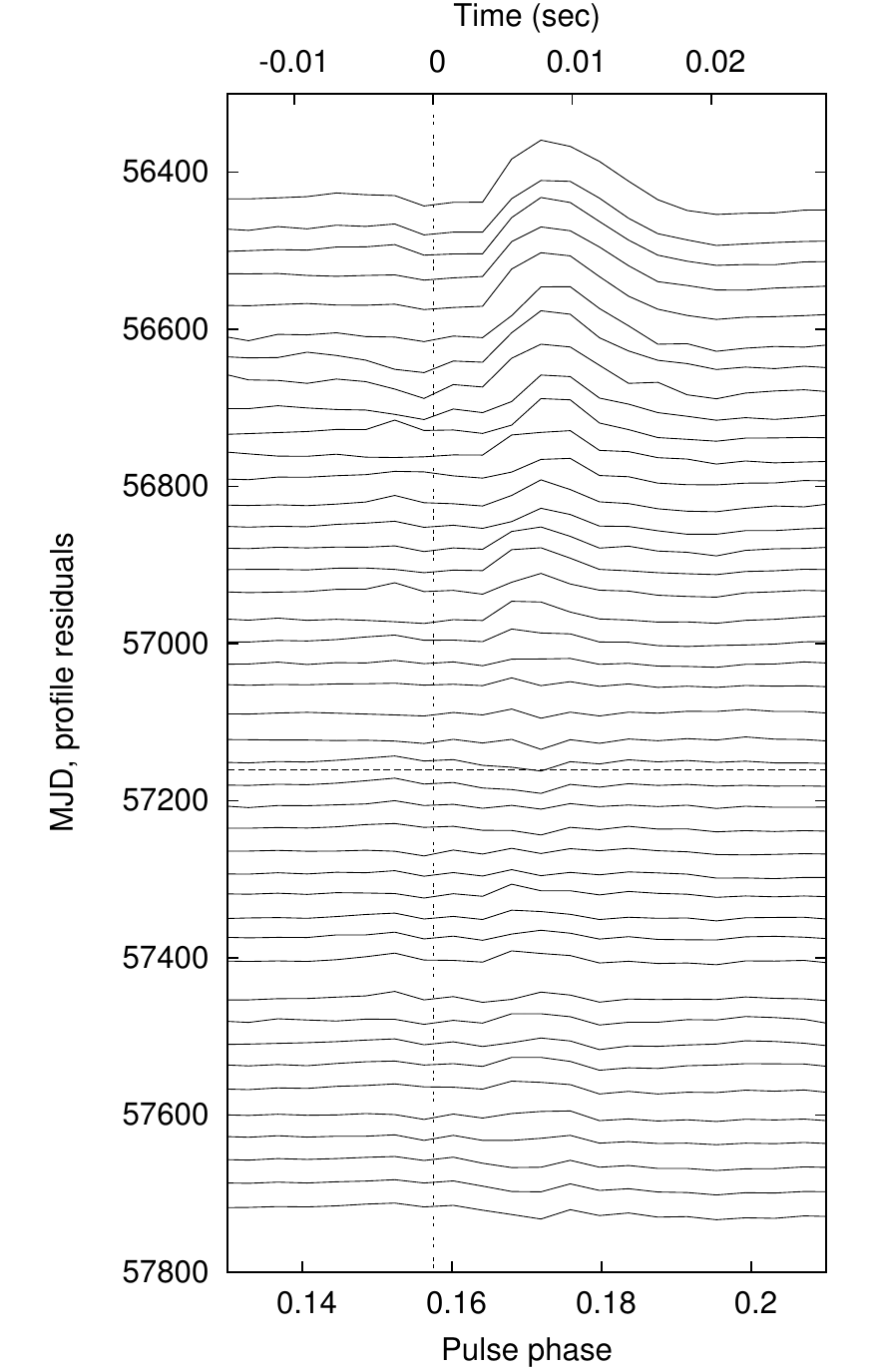}
	\caption{Pulse-shape variations for \psrj. Shown are the
      profile-shape differences between the observations and the
      standard template, which is indicated by the dashed horizontal
	 line at MJD~57161. The vertical line indicates the position of
      the peak of the pulse profile. For a clearer view, only one
      observation is plotted for every four-week interval.
      The pulse profiles used for this plot were fully
      averaged in frequency and downsampled to 256 profile bins (to
      reduce the radiometer noise in the plot).  The differences
      between observations and standard template were calculated after
      peak-normalising and
      aligning the profiles by their leading edge and peak.
	 This figure is consistent with Figure~4 of \cite{mhd+18},
	 who used a slightly extended dataset.}
	\label{waterfall}
\end{figure}

Now that the maximum impact of the profile-shape variations on our DM
measurements has been established, we investigate how the amplitude of these
profile-shape differences evolves in time. Specifically we are
concerned with the amplitude variations of the scattered power
in the trailing edge that is
visible primarily at the earlier epochs: MJD~56400--57000 at pulse
phase elongations of $\leq2\%$ from the pulse peak. This is in contrast
to the lower-lying `echo' images analysed by \citet{mhd+18}, which
have much lower amplitude (and consequently less impact on timing or
DM measurements) and lie at pulse longitudes further away from the
pulse peak ($>2\%$ of a pulse period).

\begin{figure}
	\includegraphics[width=.5\textwidth]{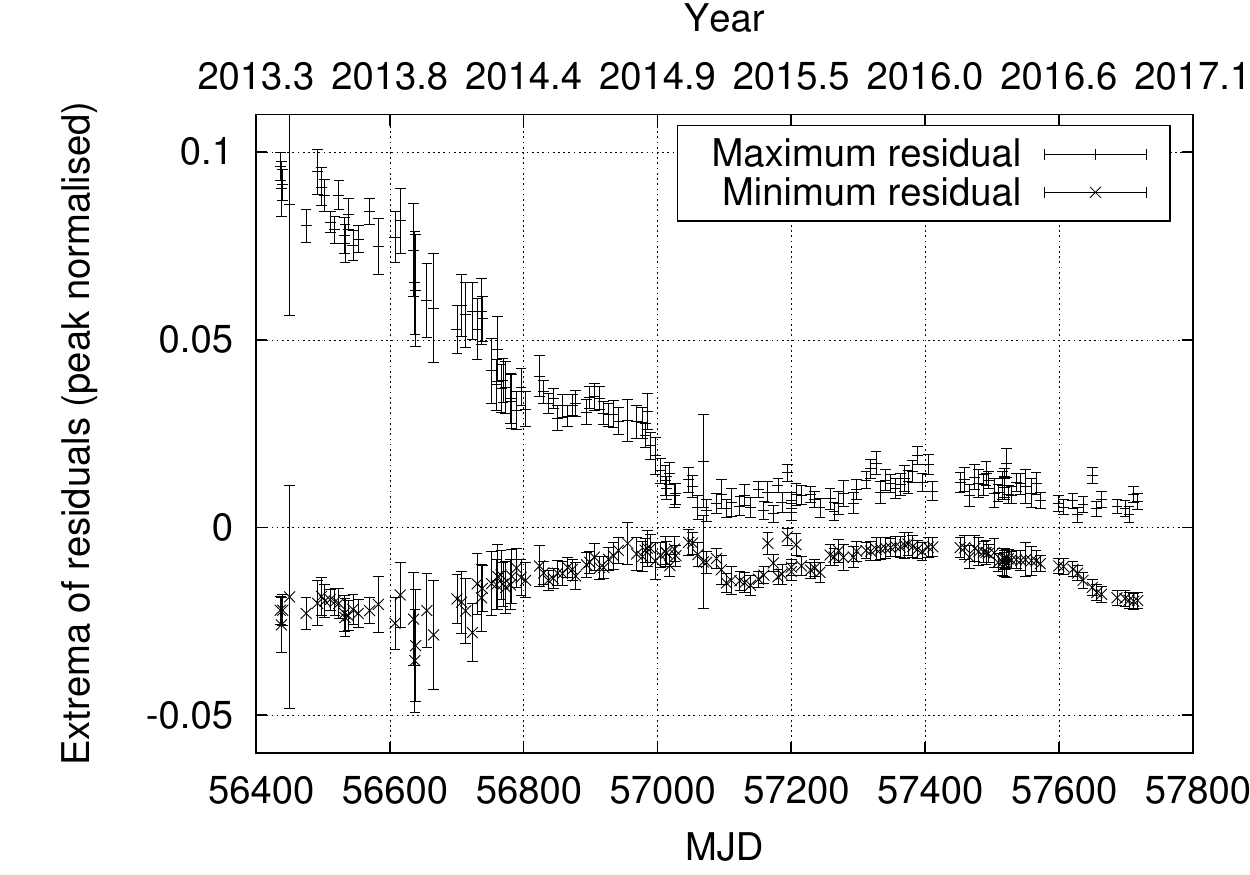}
	\caption{Maxima and minima of the profile-shape differences shown in
      Figure~\ref{waterfall}. Error bars indicate 2.88 times the off-pulse RMS
      (since for Gaussian noise there is a 1/256 chance of a measurement falling
      more than 2.88$\sigma$ away from the mean of the distribution and the
      pulse profile differences considered here consist of 256 bins).
      The profile-shape distortion due to scattering
	 is clearly visible before MJD~57000 with a residual
      amplitude that decays linearly in
      time. Between MJDs~57000 and 57600 the profile differences are not fully
      consistent with Gaussian noise, but are stable. After MJD~57600 some
      further, lower-level, variations occur.}
	\label{waterfallMinMaxRes}
\end{figure}

Figure~\ref{waterfall} shows the profile-shape differences of our
observations with respect to the template observation. This clearly
shows the excess power in the trailing edge at early epochs, with an amplitude that
decays in time. At the epoch of the template profile, the scattering
reaches a minimum and the pulse-profile is stable henceforth,
with the exception of radiometer noise and the lower-lying variations
documented by \citet{mhd+18}. To more quantitatively evaluate this
evolution, the maximum and minimum values of these profile-shape
differences are plotted as a function of time in
Figure~\ref{waterfallMinMaxRes}. Here the linear decay of the profile residuals
between the start of our data set
(MJD~56436) and MJD~57000 is clearly seen. Noting that the DM impact
of the worst-scattered profile (as quantified above) is
$1.5\times10^{-4}$\DMunits\ and that the amplitudes of any
profile-shape variations after MJD~57000 are lower by at least a
factor of five (or more), we do not expect DM corruptions at levels
beyond $10^{-4}$\DMunits\ beyond this date. For dates before MJD~57000
we consider a potential DM overestimation with amplitude $1.5\times
10^{-4}$\DMunits\ at MJD~56570 and linearly decreasing to zero by
MJD~57000. 

\subsection{Frequency-dependence of DM}
The DM time series described earlier (Figure~\ref{fig_DMvars},
Section~\ref{ssec:Timing}) was derived by carrying out a standard least-squares
fit to ToAs from the various frequency channels of any given observation, using
the \textsc{tempo2} pulsar-timing software. During these fits, the median
reduced $\chi^2$ value was 2.62, indicating that either the input ToA
uncertainties were underestimated, or that unmodelled
(i.e.\ frequency-dependent) structure was present in the post-fit timing
residuals. Non-unity reduced $\chi^2$ values are not abnormal in pulsar
timing since several reasons for inaccurate estimation of ToA errors could be
present \citep[see][for an extensive review]{vs18}.
Typically, however, these effects only cause reduced $\chi^2$
values that are lower than two \citep{vlh+16}. A more astrophysical potential
cause, which is expected to be most pronounced at low frequencies,
is chromaticity of the observed DMs.

As described in Section~\ref{ssec:Timing}, since our timing is based
on a data-derived template, we are only sensitive to differences in DM
between the observation and the template observation and hence
mere `DM chromaticity' would not be visible in our analysis. However,
if this frequency-dependent DM would be variable in time (as might be
expected given the significant changes in the overall DM), then any
non-$\nu^{-2}$ dispersive effects should be equally time-dependent.

\begin{figure*}
	\includegraphics[width=\textwidth]{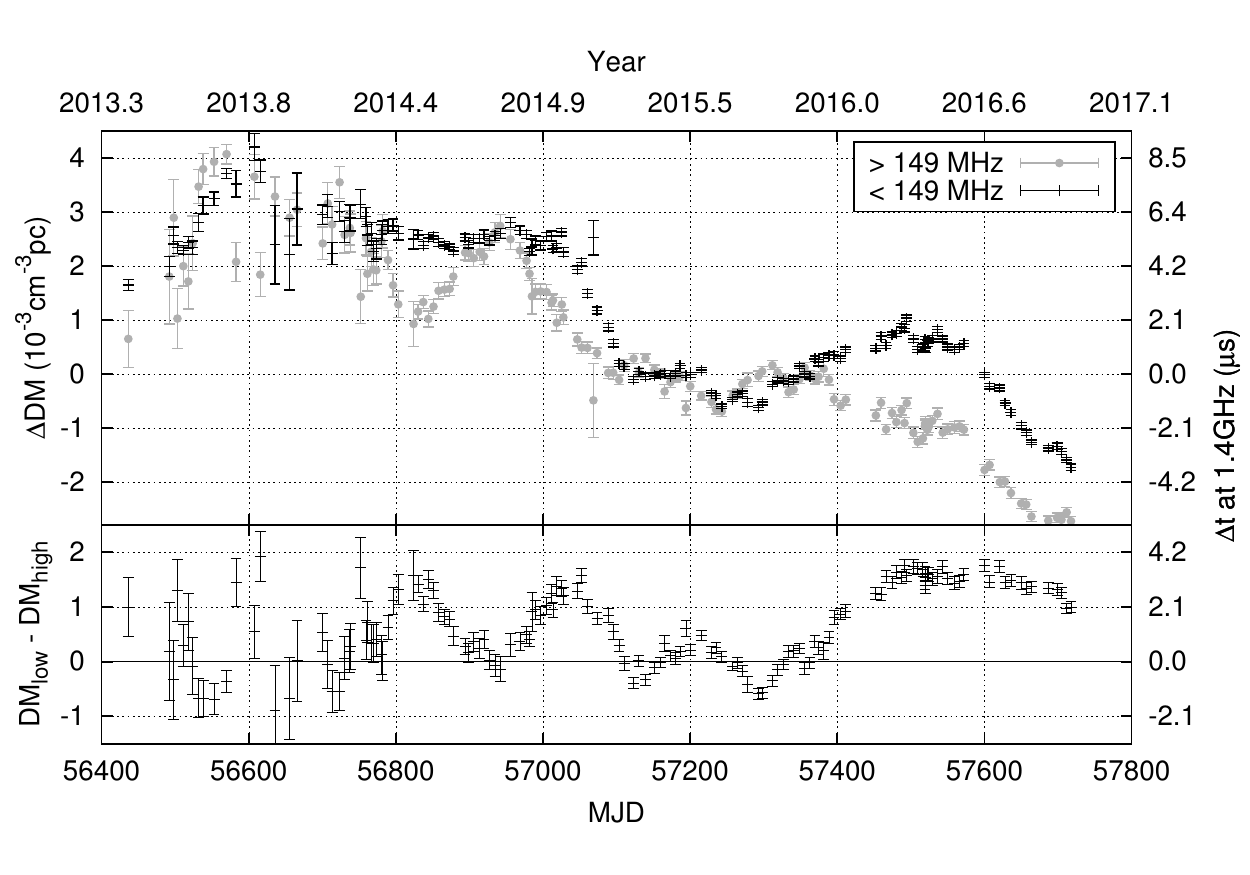}
	\caption{(top) DM time series for the upper and lower halves of the
      observing band. We note that until MJD~56524, the observing
      bandwidth was only 47\,MHz, which leads to worse DM precision
      and a smaller DM difference. (bottom) Difference between the DM
      measured in the top and bottom part of the band. By definition
      (i.e.\ through selection of the standard profile) no DM
      difference is present at MJD~57161. While the long-term DM
      gradient appears consistent between the two bands, the
      higher-frequency DM structures are shown to affect the
      high-frequency band more than the low-frequency band.
	 The second y-axis in both panels indicates the corresponding dispersive delay at
	 an observing frequency of 1.4\,GHz.}
	\label{fig_chrome}
\end{figure*}

Figure~\ref{fig_chrome} shows the time series for the DMs measured from the top
part (149--190\,MHz, centred at 169\,MHz) and bottom part (118--149\,MHz,
centred at 133\,MHz) of our band, along with the difference between these. The
frequency-dependence of these DM measurements is highly significant and exceeds
the maximal impact of the profile-shape variability, as quantified in
Section~\ref{DMvars_test}, by an order of magnitude.

\subsection{Structure function analysis}
Given the strong and significant DM variations we detected -- as well as the
frequency-dependence of these measurements -- it is worthwhile to
evaluate the overall structure of the IISM towards this pulsar in
order to verify whether the DM variability could be explained by standard IISM turbulence or
not. To this end, we compute the structure functions of the DM
variations shown in Figures~\ref{fig_DMvars} and \ref{fig_chrome} and compare these to a
Kolmogorov turbulence density spectrum, which is known to usually be a
good approximation for the IISM density spectrum in general
\citep{ars95,kcs+13}, although deviations have been reported \citep[see][for a review]{gup00}.
The structure function at a given time lag $\tau$ is derived from the DM
time series using the following equation:
\begin{equation}\label{eq_DDM}
	D_{\rm{DM}}(\tau) = <[\rm{DM}(t+\tau)-\rm{DM}(t)]^2>
\end{equation}
using a weighted mean (i.e.\ the $\Delta$DM values were weighted by
$1 / (\sigma_{\text{DM}(t+\tau)}^2+\sigma_{\text{DM}(t)}^2)$).
The uncertainties of $D_{\rm{DM}}(\tau)$ are derived through Monte-Carlo simulations, by
varying the DM time series according to the DM measurement
uncertainties and thereby identifying the 68\% confidence intervals of
$D_{\rm DM}(\tau)$ over 10,000 simulations.

\begin{figure}
  \includegraphics[width=.5\textwidth]{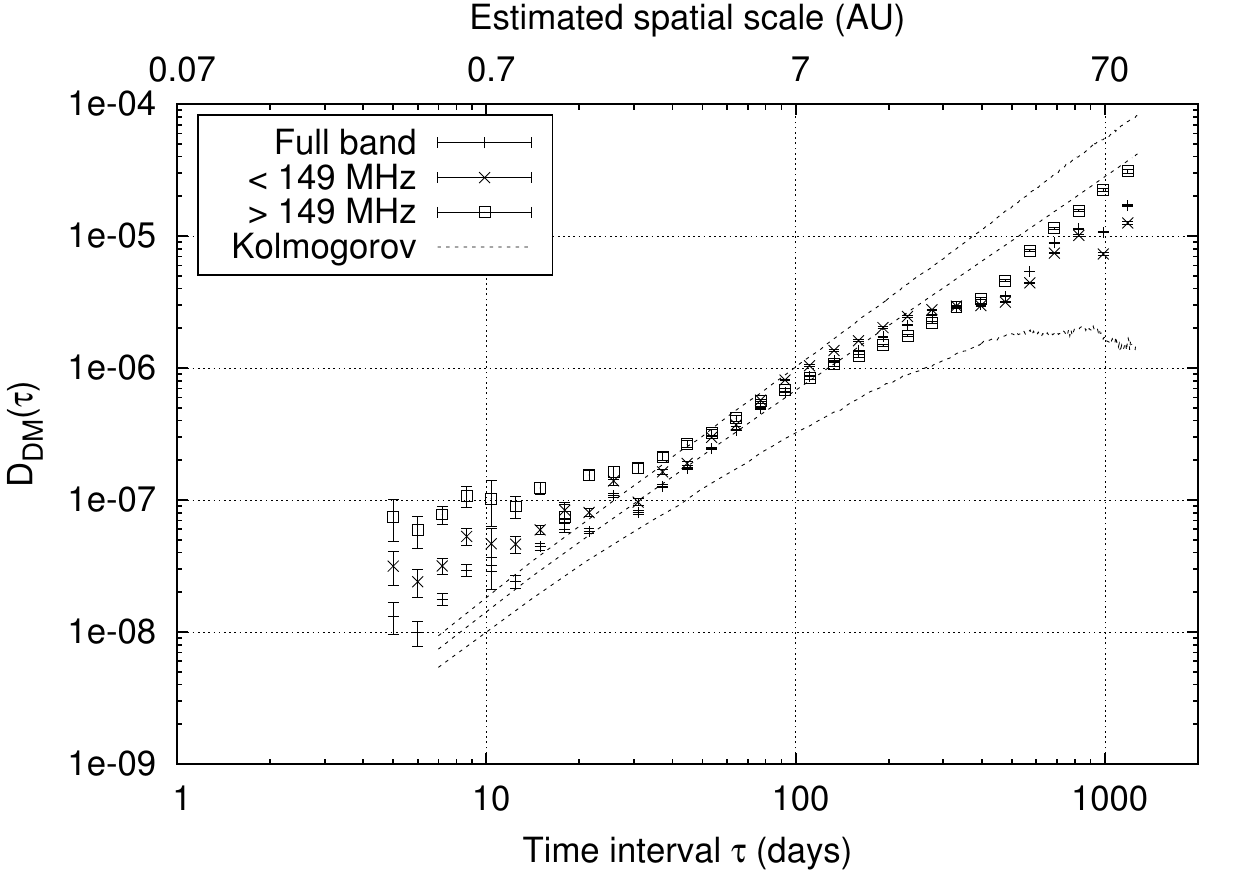}
	\caption{Structure functions of the DM time series from
      Figures~\ref{fig_DMvars} and \ref{fig_chrome}. The spatial scale
      is estimated for turbulence half-way to the pulsar, using the
      proper motion published by \citet{mhd+18}.
	The Kolmogorov turbulence model is characterised by the structure functions
	of simulated DM time series derived from a power spectrum with spectral index 5/3.
	The dotted lines represent the sample average and 1$\sigma$-contours of 1000 iterations.
	The amplitude of the Kolmogorov model was scaled to fit the first 200 days of the data,
	because the structure function becomes very uncertain at the longest lags.
	}
	\label{fig_struct}
\end{figure}

Figure~\ref{fig_struct} shows the structure function of the DM time series presented in
Figures~\ref{fig_DMvars} and \ref{fig_chrome}. Due to the high observing cadence with the GLOW stations,
the shortest lags we sample in our structure function go down to a few days.
Previously published structure functions of DM time series sampled time lags down to
10s of days \citep{yhc+07,jml+17} or about 100 days \citep{kcs+13},
so with our dataset we extend the range at which Kolmogorov turbulence has been tested
in DM time series by one order of magnitude.

We find that the structure function of the DM time series agrees extremely well
with a Kolmogorov spectrum.
This conformity seems to corroborate the findings by \citet{pk12,pl14}
that ESEs may actually be part of the larger-scale IISM turbulent structure
rather than separate, localised, events.

Using Eq.~28 of \cite{lcc+16} for Kolmogorov turbulence, neglecting Earth's motion and assuming a constant amplitude of the electron density power-law spectrum ($C_N^2$) along the ling of sight, the latter can be directly related to the amplitude of the DM structure function:
\begin{equation}
	D_\text{DM}(\tau) = C_N^2 \cdot 88.3\cdot\frac{3}{8}\ z_p^{8/3}\ \mu^{5/3}\ \tau^{5/3},
\end{equation}
with the proper motion $\mu$ and the pulsar distance $z_p$.
From the Kolmogorov power-law fit to the structure function for time scales up to 200 days, we get
\begin{equation}
	D_\text{DM}(\tau) = 3.1\cdot10^{-10}\,\text{cm}^{-6}\,\text{pc}^2
					\left(\frac{\tau}{1\text{d}}\right)^{5/3}.
\end{equation}
With a proper motion of 22.2\,mas/yr \citep[see][]{mhd+18}
and a pulsar distance of 2200\,pc \citep{cl02}\footnote{
	\url{https://www.nrl.navy.mil/rsd/RORF/ne2001/}.
	Recently, \cite{ymw17} published a new Galactic electron density
	model (\url{http://119.78.162.254/dmodel/}), which gives a
	consistent distance estimate of 2.4\,kpc.
},
we get $C_N^2 = 0.9\cdot10^{-3}m^{-6.67}$. This is in close agreement with the findings of \cite{ars95}, that $C_N^2 = 10^{-3}m^{-6.67}$ fits a huge range of spatial scales and thus supports the idea that the observed variability is part of the general IISM turbulence rather than a stand-alone ESE.

\section{Discussion}
\label{sec_discussion}
\subsection{Origin of the DM variability}
\label{sec_cloud_model}
Both the amplitude and shape of the DM variations shown in
Figure~\ref{fig_DMvars} are reminiscent of ESEs presented elsewhere in the
literature \citep[e.g.][]{cbl+93,cks+15} and interpreted as individual lenses of ionised matter.
In addition to this, the scattering events seen in this pulsar and discussed in our
companion paper \citep{mhd+18} are consistent with a number of contained,
refractive lenses near the line of sight. Nevertheless, the structure function
(Figure~\ref{fig_struct}) is fully consistent with a Kolmogorov spectrum. In
the following, we will consider a simplistic model based on three individual,
spherical interstellar lenses in order to compare the required lens sizes
and densities to those previously published for ESEs.\footnote{We cannot analyse
  potential correlations between scintillation and DM variability like
  \citet{cks+15} did because the scintillation bandwidth is smaller than our
  frequency resolution.} We presume these lenses to be the cause of the DM
excess around MJDs~56600 and 57000. The steep drop-off in DM towards the end of
our data set will be disregarded as further monitoring of this pulsar's DM time
series is required for this event. (Particularly the apparently third ESE which
seems to only affect the lower-frequency part of our data deserves further
analysis and continued observations.)

In order to identify the presumed ESEs clearly, we define a baseline
DM level of 43.482 \DMunits, which is the
average DM value from MJD~57100 to MJD~57200.
This value is chosen as
reference because it describes the only time window in our data where
no variations in DM are observed. It is also used as the reference
value for all plots. The reference observation against which the
timing was performed was also selected from within this MJD range.
We furthermore only consider the DM peak near MJD~57000
because this is the most easily identified and the differences with
the other potential ESEs are well within the uncertainties of our
model, so similar results can be considered to hold for all three
presumed clouds.

We will now model the aforementioned DM peak, assuming a spherical,
homogeneous cloud of ionised gas to be causing it.
The time from the maximum to the end of the peak is 150\, days, so the
total duration of the cloud passage is estimated to be 300\,days. Using a proper
motion of 22.2\,mas/yr \citep[see][]{mhd+18}, the angular size of the cloud can be
calculated as $\theta = 18\,\text{mas}$. To estimate the physical
size, the distance to the cloud is needed. This distance can only be
estimated, but it has to be lower than the distance to the pulsar,
which is estimated to be 2.2\,kpc \citep{cl02}. Combining the angular size
of the ESE and the distance to the pulsar, the maximum size of the
cloud (if it was directly in front of the pulsar) is 40\,AU.

\begin{figure}
	\centering
	\includegraphics[width=.5\textwidth]{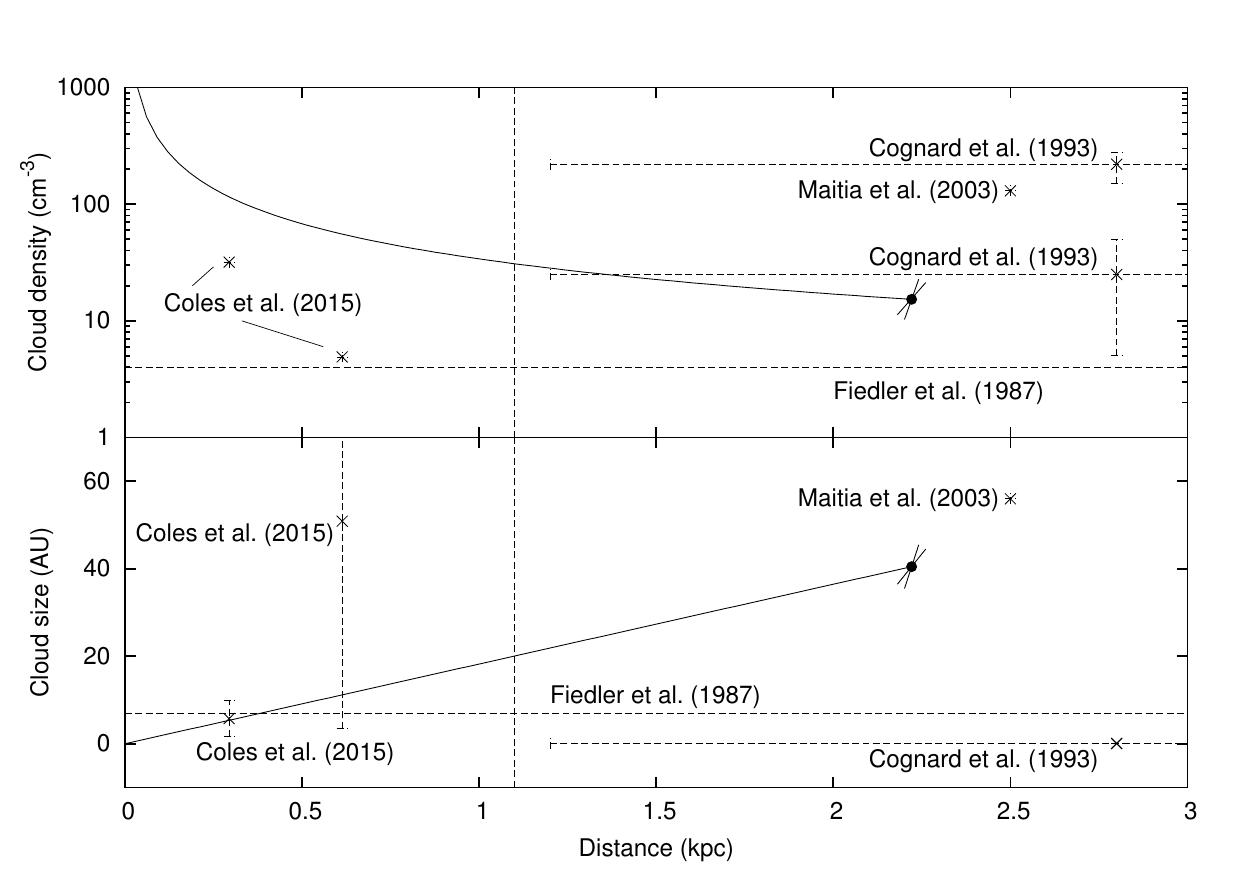}
	\caption{Estimated cloud size and electron density for the second
      component of the ESE depending on its distance to the Earth,
      represented as solid lines in the plot. Horizontal lines
      represent literature values without distance estimates.
	 The vertical line represents the distance estimate for the structures
	 that cause the scattering `echoes' from our companion paper \citep{mhd+18}.
	 The cloud size estimates of
      \cite{cbl+93} for the two clouds forming the ESE
      are very similar (0.050\,AU and 0.094\,AU) and thus
      indistinguishable in the plot. The error bars on the cloud sizes
      of \citet{cks+15} represent the non-spherical character of the
      clouds and not the actual uncertainty of the cloud size.
	 }
	\label{fig_cloud_size}
\end{figure}

Given the assumption of a spherical object, the maximum path length
through the cloud is equal to its lateral extent, and from this path
length and the maximum DM increase one can calculate the average
excess electron density in the cloud. From the upper limit on the
cloud size follows a lower limit on the average extra electron
density, which is $15\,$cm$^{-3}$. Compared to a typical electron density
in the Warm Ionised Medium of $0.1 - 0.5\,\text{cm}^{-3}$ \citep{hs13}, our model
suggests an overdensity of about two orders of magnitude.
The solid lines in Figure~\ref{fig_cloud_size} show
the estimated cloud size and its electron density depending on its
distance to the Earth. Some values from the literature of similar
estimates are added for comparison: the first ESE observed by
\citet[without distance estimate]{fdjh87}, an ESE observed in flux
density and timing residuals modelled as two clouds of identical size by \citet{cbl+93},
the three-year-long ESE observed in flux density by \citet[without
  uncertainties]{mlc03}, and the models of two ESEs observed by
\citet{cks+15}. These latter two ESEs were rescaled to be compatible
with pulsar distances based on the parallax measurements of
\citet{nbb+14} and \citet{rhc+16}, corrected for the Lutz-Kelker bias
following the analysis by \citet{vwc+12} as corrected by
\citet{ivc16}.

When comparing the size and density estimates to typical values from the
literature, this particular ESE could be anywhere along our line of
sight without falling out of the sample, as the densities and sizes in
the literature span the entire range of possible values (with an
exception for clouds that are very close to the Earth as the implied
density would become unrealistically high).
In particular a cloud roughly halfway to the pulsar
\citep[in agreement with][]{mhd+18} with a size of
about 20\,AU and a density of a few tens of electrons per cm$^3$
would be highly comparable to previously observed ESEs.
All calculations for this
simple model are only very rough estimates, as it is impossible to
disentangle the different components of the DM time series and to find
the correct DM baseline. It is, for example, highly likely that there are
multiple clouds of various sizes, which overlap \citep[see also the
  discussion by][]{mhd+18}.  The steepest DM decrease, which starts
around MJD~57000, can be interpreted as the edge of one cloud
and lasts about 75\,days. Estimating the cloud to be located half-way
to the pulsar, this edge is about 5\,AU thick. This is roughly of the
order of what \cite{bmg+10} found for elongated filaments, so it could
be the case that the filaments they see are actually the edges of
ionised clouds \citep[see also][]{pl14,lpm+16}.

\subsection{Origin of the DM frequency-dependence}
In the following, we will assume that the observed DM difference was caused by deviations
from the dispersion law (Eq.~\ref{eq:t_DM}) due to invalidity
of the assumptions that $\nu \gg \nu_p$ or $\nu \gg \nu_c$.

The observed variability of $\Delta \text{DM}$ of the order of $10^{-3}$\,\DMunits{}
between our two bands centred at 133\,MHz and 169\,MHz
corresponds to an additional time delay of the order of $\sim200\mu s$,
which is smaller than the total dispersive delay by a factor of about 5e-5, so the variability
of the sum of $T_1$ and $T_2$ would have to be of the order of 5e-5 as well (see Eq.~\ref{eq:tzd68}) if we assume a homogeneous medium along the line of sight.
To relate $T_1$ to an electron density, we insert Eq.~\ref{eq:nu_p} into Eq.~\ref{eq:T1} and solve for $n_e$. To relate $T_2$ to a magnetic field strength along the line of sight, we similarly insert Eq.~\ref{eq:nu_c} into Eq.~\ref{eq:T2} and solve for $B_\parallel = B \cos \gamma$:
\begin{equation}
	\frac{n_e}{\text{cm}^{-3}} =
	\frac{1}{(8.98\,\text{kHz})^2}\cdot \frac{4}{3} T_1 \frac{\nu_1^2 \nu_2^2}{\nu_1^2 + \nu_2^2}
	= 1.81\cdot10^{8}\ T_1
\end{equation}
\begin{equation}
	\frac{B_\parallel}{1\text{G}} =
	\frac{1}{2.80\,\text{MHz}}\cdot \frac{1}{2} T_2	\nu_1\nu_2 \frac{\nu_2^2-\nu_1^2}{\nu_2^3-\nu_1^3}
	= 1.76\cdot10^{1}\ T_2
\end{equation}
$T_1$ or $T_2$ of the order of 5e-5 would imply variations of the electron density of the order of $10^{4}\,\text{cm}^{-3}$, or variations of the magnetic field along the line of sight of the order of $10^{-3}\,\text{G}$.
This is equivalent to the DM and RM varying by many orders of magnitude of their actual value.
As shown in this paper, the DM only changes by fractions of
its total value. A detailed analysis of the RMs in these data is in
progress and will be published in due course; but any RM variations
towards this source are minimal and would not satisfy this scenario, either.

As DM chromaticity is usually not observed, a localised structure as the origin seems far more likely. If a structure corresponding to a DM excess of the order of $10^{-2}$\,\DMunits{} was causing the observed variations of the frequency-dependence (as expected from the DM time series in Fig.~\ref{fig_DMvars}), this would require variations in $T_1 + T_2$ of the order of 0.2, implying electron density variations of the order of $4\cdot10^{7}\,\text{cm}^{-3}$, or variations of the magnetic field strength of the order of 3\,G. To only contribute $10^{-2}$\,\DMunits{} to the DM, the structures would need to be 5e-5\,AU thick, which is in strong contradiction with the long duration of these variations.
Specifically, a 5e-5 AU-thick region with an electron-density excess
of the order $10^{7}\,\text{cm}^{-3}$ could be generated by a chance alignment of a star
with the line of sight, but given the rapid spatial motion of the line
of sight, such an alignment would pass quickly. \cite{mhd+18}
also discussed the potential impact of stellar winds on the
observations of this pulsar, but in that scenario the offset of the
relevant star to the line of sight is too far to allow the mechanism
proposed by \cite{tzd68} as an explanation for the
frequency dependence of the DM since the size and density of the
stellar-wind bubble along the line of sight would not match the
predictions derived above.

As these scenarios are highly unrealistic,
we conclude that deviations from the dispersion law as described by \cite{tzd68}
are not the cause of the frequency-dependent DMs we detected.
We therefore favour the explanation from \cite{css16},
that refractive effects lead to a frequency-dependence
of the medium the pulsar radiation passes through.
While the DM time series at 133\,MHz does on the
whole look smoother than the variations at 169\,MHz \citep[as expected from the
  analysis by][]{css16}, at a few epochs (e.g.\ around MJD~57100 and shortly after
MJD~57200) the lower frequencies show more dramatic DM trends,
which may be in tension with the theoretical expectations of this model.

\subsection{Consequences for high-precision pulsar timing}
Variations in the DM that cannot be accurately and precisely measured and
modelled are a problem for pulsar timing, as they add a time-dependent extra
delay to the ToAs. If the DM variations we reported here were to occur along the
line of sight to pulsars used in high-precision timing experiments, they would
corrupt astrophysically relevant parameters and would significantly reduce
sensitivity to interesting signals \citep[see, e.g.][]{yhc+07}
if their impact on the data was not removed by, for example, measuring the DM at
every epoch and correcting for it. This correction can only be done with
simultaneous multi-frequency or low-frequency data, which are not always
available \citep[see e.g.\ the first IPTA release,][]{vlh+16}. In the potential ESE
discussed in this paper, the maximum difference in the extra time-delay across
the entire dataset (caused by a DM difference of $6\times10^{-3}\,$\DMunits) at
the commonly used 21-cm wavelength would imply structures of the order of
$13\,\mu\text{s}$ in the timing residuals. This is well above the precision needed for high-precision pulsar timing experiments, which require sub-microsecond precision
\citep[see, e.g.][]{jhlm05}. The ToA difference across
a 250-MHz bandwidth centred at 1.4\,GHz due to the extra dispersion is
$2\,\mu\text{s}$, so the ToA precision has to be substantially better to
properly measure and correct the impact the ESE has on the data \citep[we note
  that the median timing precision in the IPTA is currently
  2.5\,$\mu$s,][]{vlh+16}. As shown by \citet[Eq.~12]{lbj+14}, the precision of
this correction would however be about an order of magnitude worse than the ToA
precision and averaging the DM values to increase precision is typically not a
valid solution because of the usually low sampling rate \citep[see][and
  references therein]{vlh+16} and possible short time scales of the
variations. Thus, correcting high-frequency observations with DM values measured
from that same observation would not suffice to correct for DM variations
similar to the ones presented in this paper.

While low-frequency data are very useful in computing highly detailed
measurements of DM variations, the chromaticity we have presented does
cause concern as it may imply a mismatch between the DM values
observed at lower frequencies and those observed at higher
frequencies. We note, however, that chromaticity mostly perturbs DM
variations on short timescales, whereas the long-term DM trends tend
to show reasonable levels of agreement \citep[as suggested by][and
  confirmed by our analysis, see the bottom panel of
  Figure~\ref{fig_chrome}]{css16}. Since the spatial electron-density spectrum
in our Galaxy (and therefore the spectrum of DM variations) has more
power at lower frequencies \citep{ars95}, it is particularly
the longer-term DM variations that require correction in
high-frequency pulsar-timing data, which implies the impact of
chromaticity may be limited. Further studies of chromatic DM
variations in pulsars like \psrj{} -- particularly studies that
extend our present analysis to include a wider range of observing
frequencies -- should allow more conclusive answers to these
questions.

\section{Conclusions}\label{sec_concl}
We have presented strong and rapid DM variations along the line of sight towards \psrj,
which have a similar amplitude as the variations commonly seen in millisecond pulsars
\citep[see, e.g.][]{kcs+13}.
The variations we reported may be caused by a group of interstellar clouds typically referred to as
ESEs. The ESEs in this paper would be some of the longest and most persistent
observed to date. This is also the first time an ESE is observed in electron
density and scattering at the same time, although the scattering may
well be caused by different IISM structures that have only a minimal
impact on the DM \citep[for the quantitative analysis of the scattering, see the companion paper by][]{mhd+18}.

Our frequent observations with the international LOFAR stations allow detailed
and highly precise monitoring of the DM time evolution.
The high measurement precision makes us sensitive to details, which complicates any
efforts to provide an accurate model for the underlying IISM structures.
Additionally, there is a large number of unknown parameters like the distance
or proper motion of the IISM structures and the DM baseline level. The simple spherical
model discussed in this paper does not provide definite values, but does provide
limits on the potential ESE's electron density and size, which are of the same
order of magnitude as previous results for ESEs. Further observations of ESEs
will help to improve the constraints on the size, because the lack of knowledge
of the distance to the IISM structures is less of an issue when a larger sample of observations
is given, assuming a homogeneous distribution of ESEs in the Galaxy.  We
furthermore point out that the variations presented here could well be part of a
uniform spectrum rather than separate, distinct, structures.

Finally, we have presented the first observational evidence for
frequency-dependent DMs and have confirmed that the long-term
DM trends are consistent across the frequencies we probed, whereas the
shorter-term DM structure is highly chromatic. This bodes well for
efforts to apply low-frequency DM time series as corrections to
high-frequency pulsar-timing data, although further study across a
wider range of frequencies should be undertaken to quantify
any potential corruptions such corrections would cause.

\begin{acknowledgements}
JYD acknowledges financial support during part of this work from a
Br\"uckenstipendium granted by Bielefeld University.
SO acknowledges the support from the Alexander von Humboldt Foundation and the Australian Research Council grant Laureate Fellowship FL150100148.
JWTH and DM acknowledge support from the European
Research Council under the European Union's Seventh Framework
Programme (FP/2007-2013) / ERC Grant Agreement nr. 337062
(`DRAGNET').
This paper is based on data from the German LOng-Wavelength (GLOW) array, which
is part of the International LOFAR Telescope (ILT) which is designed and built
by ASTRON \citep{vwg+13}. Specifically,
we used the Effelsberg (DE601) station funded by the Max-Planck-Gesellschaft;
the Tautenburg (DE603) station funded by the State of Thuringia and supported by
the European Union (EFRE) and the Federal Ministry of Education and Research
(BMBF) Verbundforschung project D-LOFAR I (grant number 05A08ST1);
and the J\"ulich (DE605) station supported by the BMBF Verbundforschung project
D-LOFAR I (grant number 05A08LJ1). 
The observations of the German LOFAR stations were carried out in the
stand-alone GLOW mode which is technically operated and supported by the
Max-Planck-Institut f\"ur Radioastronomie, the Forschungszentrum J\"ulich,
Bielefeld University, by BMBF Verbundforschung project D-LOFAR III (grant
number 05A14PBA) and by the states of Nordrhein-Westfalen and Hamburg.
The observations during this project were made during station-owners time as
well as during ILT time allocated under project codes LC0\_014, LC1\_048,
LC2\_011, LC3\_029, LC4\_025 and LT5\_001.
This project also benefited greatly from the ILT Core observations
that were included in the Michilli et al. (2018) paper and which
guided some of the analysis and interpretation in this paper. Those
observations were carried out under ILT time allocation codes
LC0\_011, LC1\_027, LC2\_010, LT3\_001, LC4\_004 and LT5\_003.
Plots and basic model fits were created with \textsc{gnuplot}.
\end{acknowledgements}

\bibliographystyle{aa}
\bibliography{journals,psrrefs,modrefs,crossrefs}

\end{document}